# Nonlinear Hall effect in bilayer $WTe_2$ induced by strong laser fields

*M. Umar Farooq[1]†, Arqum Hashmi[2]†, Mizuki Tani[3], Kazuhiro Yabana[4], Kenichi L. Ishikawa[2]*, Li Huang[1,5]*, and Tomohito Otobe[3]**

[1] *Department of Physics, State Key Laboratory of Quantum Functional Materials, and Guangdong Basic Research Center of Excellence for Quantum Science, Southern University of Science and Technology (SUSTech), Shenzhen, 518055, China*

[2] *Department of Nuclear Engineering and Management, Graduate School of Engineering, The University of Tokyo, 7-3-1 Hongo, Bunkyo-ku, Tokyo 113-8656, Japan*

[3] *Kansai Institute for Photon Science, National Institutes for Quantum Science and Technology (QST), Kyoto 619-0215, Japan*

[4] *Center for Computational Sciences, University of Tsukuba, Tsukuba 305-8577, Japan*

[5] *Quantum Science Center of Guangdong-Hong Kong-Macao Greater Bay Area (Guangdong), Shenzhen 518045, China*

**AUTHOR INFORMATION**

**† These authors contribute equally**

**Corresponding Author**

*E-mail: ishiken@n.t.u-tokyo.ac.jp, huangl@sustech.edu.cn, otobe.tomohito@qst.go.jp

## ABSTRACT

The study of the interplay between the geometric nature of Bloch electrons and transverse responses under a strong field offers new opportunities for optoelectronic applications. Robust nonlinear anomalous Hall phenomena can originate in inversion asymmetric crystals with avoided band crossings due to large Berry curvature peaks. Here, we investigated the relationship between nonlinear Hall effects and nonlinear optical responses of bilayer $WTe_2$, ranging from perturbative to nonperturbative regimes. The high-harmonic generation (HHG) spectra under a strong field clearly exhibit a plateau and energy cutoffs, i.e., an unambiguous signature of nonperturbativeness, for both longitudinal and anomalous Hall (transverse) currents, with the latter being due to the large interband Berry curvature. For the longitudinal harmonics, the intraband contributions increase with intensity, resulting in a complex interplay between interband polarization and intraband motions. If we take a comprehensive *all-band* perspective enabled by time-dependent density functional calculations and define the intraband current as the diagonal part of the total current density matrix obtained through the projection operation, the anomalous Hall responses are attributed to the interband processes, even in the nonperturbative regime, though the decomposition into the intra- and interband components is arbitrary rather than physically essential. Thus, Hall HHG can be crucial to understand the carrier dynamics. Our findings suggest that HHG associated with the ultrafast strong-field driven electron dynamics holds immense potential for exploring the nonlinear high Hall responses in noncentrosymmetric semimetals.

# I. INTRODUCTION

The transverse electric responses, commonly referred to as the Hall effects, are at the center of various condensed matter research areas, including dissipation-less topological edge states, magnetic topology, spin-orbit coupling effect, and disorder scatterings [1–6]. Within the linear limits, electronic transverse responses originate from non-zero Berry curvature and are subject to broken time-reversal symmetry [7]. In the nonlinear domain, symmetry constraints change, and structural inversion asymmetry can lead to nonlinear second-order anomalous Hall effects [8]. While the static nonlinear Hall effects can be observed, experimental studies have focused primarily on the quantum doubling of transverse frequencies in Weyl and Dirac semimetals [9,10]. The physical origin of this nonlinear transverse current is attributed to the pronounced Berry curvature dipolar peaks in the momentum space, which result from the Dirac-like band dispersions.

The existing experimental efforts regarding nonlinear Hall currents have mainly focused on low frequencies, where interband transitions are avoided, and the conductivity of the system rarely shows frequency dependence [11,12]. Simultaneously, theoretical investigations are predominantly based on the geometric nature of Bloch electron and perturbation theory emphasizing intraband contributions [13–15]. On the other hand, high harmonic generation (HHG) studies have mainly focused on semiconductors [16] and Weyl semimetal [17], tentatively suggesting that the anomalous transverse even order current responses arise from the intraband motion and Berry curvature within the single-band picture. However, interband coherence can also contribute to nonlinear anomalous Hall currents, which intrinsically can be linked to the interband Berry curvature dipole or multipoles depending on the order of nonlinearity [13,18]. Previous studies based on the two-band model have also discussed the impact of time-reversal symmetry breaking, which results in odd-order Hall harmonics, as well as the relationship between HHG and the nature of Weyl nodes have been explored [19–22]. However, in the strong field regime, interband resonance involving bands away from the fermi level becomes important and cannot be adequately described by such simplified models [23].

Moreover, Dirac semimetals are predicted to exhibit fascinating field-induced electron dynamics owing to their ultrafast optical response and weak screening [24]. In the weak-field (perturbative) regime, the electronic current response and excitations can be elucidated by quantized-photon excitations [25]. On the other hand, under a strong field, the electron dynamics shift towards a light-field-driven regime, where tunneling mechanisms dictate electron motions, rendering perturbative explanations inadequate. In Dirac semimetals such as graphene, this regime involves interband transitions coupled with intraband oscillations within a Landau–Zener transition framework [26,27]. Therefore, with the change in electric field strength, the nonlinear phenomena can originate from various mechanisms, including Bloch oscillations, intraband motions, and electron-hole recombination. However, the contributions of different processes to the higher-order Hall have not yet been investigated. Additionally, exploring the interplay between longitudinal and transverse currents in different regimes is crucial.

Due to its inherent spatial asymmetry and weaker charge carrier screening, the bilayer T'-$WTe_2$ semimetal offers a promising platform for investigating Dirac electron dynamics [19,28,29]. Furthermore, the observations of the nonlinear Hall effect in bilayer T'-$WTe_2$ [8] driven by its strong Berry curvature dipole, presents a unique opportunity to investigate the interaction between strong fields and transverse responses. Based on time-dependent density functional theory (TDDFT) calculations, we show that the high-harmonic spectra clearly exhibit a plateau for both longitudinal and anomalous Hall (transverse) components, which is an unambiguous evidence of a nonperturbative response. Furthermore, the effect of intensity and frequency dependence, including cutoff inter/intraband effects on Hall dynamics are discussed in detail. We also study the role of distinctive excitation mechanisms, i.e., tunneling, the interplay of intraband and interband, and single/multiphoton absorption, on the high-order anomalous Hall (transverse) current concerning the Dirac points. Remarkably, the results display a paradigm shift in carrier dynamics from purely interband to increasing intraband contributions as the field strength increases. Our analyses show that, even in the nonperturbative regime, the physical origin of Hall harmonics stems from interband processes, as observed from our comprehensive all-band perspective. While this may initially appear distinct, it is fundamentally consistent with previously reported interpretations from single-band perspectives. Thus, our study reveals the intricate interplay between ultrafast carrier dynamics and nonlinear Hall responses.

## II. THEORETICAL FORMALISM AND COMPUTATIONAL DETAILS

The TDDFT calculations are performed within the framework of SALMON (Scalable Ab-initio Light-Matter simulator for Optics and Nanoscience) [30–32]. The time-dependent Kohn-Sham (TDKS) equation for the Kohn-Sham orbital $\psi(r,t)$, in velocity gauge is described as [24],

$$i\hbar \frac{\partial}{\partial t}\psi_i(r,t) = \left[\frac{1}{2m}\left(-i\hbar\nabla + \frac{1}{c}A(t)\right)^2 - eV(r,t) + v_{ion}(t) + v_{XC}(r,t)\right]\psi_i(r,t) \quad (1)$$

where $A(t)$ is the vector potential, $V(r,t)$ is a scalar potential that contains the Hartree potential and the local part of the ionic potential. The $v_{ion}$ and the $v_{XC}$ are the nonlocal part of ionic pseudopotential and the exchange-correlation potential, respectively.

The electric current density $J(t)$ is calculated by solving Eq. (1) in the time domain [33]. The macroscopic electric current density $J(t)$, which is the spatial average of the microscopic electronic current density over the unit cell volume is given as follows,

$$J(t) = \frac{-e}{m} \int_{\Omega} \frac{dr}{\Omega} \sum_i \psi_i(r,t)^* \left(-i\hbar\nabla + \frac{e}{c} A(t)\right) \psi_i(r,t) + \delta J(t) \tag{2}$$

where $\Omega$ is the volume of the unit cell while $\delta J(t)$ is the current caused by the nonlocality of the pseudopotential [34,35]. $J(t)$ expresses the macroscopic current density in three-dimensional space. However, note that the thickness is regarded as infinitesimally small for two-dimensional (2D) materials in the macroscopic sense the 2D current density can be defined as $J_{2D}(t)$ by integrating over z-direction (perpendicular to the layer) and averaging over $xy$-plane (plane of the 2D material),

$$J_{2D}(t) = \int_{-\infty}^{\infty} dz \int_{\Omega} \frac{dxdy}{\Omega} J_{3D}(x,y,z,t) \tag{3}$$

where Ω is the 2D unit-cell area of the 2D material.

The excited electron population is defined as

$$\rho_{\mathbf{k}}(t) = \sum_{occ} \left| \int_{\Omega} d^3r \sum_c \ \psi^*_{v,k}(r,t)\, \psi^{\mathrm{G}S}_{c,k+\frac{e}{\hbar c}\mathrm{A}(t)}(r) \right|^2 \tag{4}$$

where $v$ and $c$ are the indices for the valence and conduction bands, respectively, and $\psi_k^{\mathrm{G}S}(r) = \psi_k(r, t=0)$ is the Bloch orbital at the ground state. To split the intraband current component, the Kohn-Sham orbital projected onto the adiabatic Houston basis $\varphi_i(k + A(t))$ which is the eigenstate of time-dependent Hamiltonian as

$$\psi_i(t) \approx \sum_{i'} \langle \varphi_{i'} | \psi_i(t) \rangle \, \varphi_{i'} \equiv \sum_{i'} \mathrm{C}_{i'i}(t)\, \varphi_{i'} \tag{5}$$

where the $i'$ runs on both occupied and unoccupied orbitals. After getting the complete set of $\varphi_{i'}$ including valence and conduction bands, the intraband currents can be calculated as

$$J_{\mathrm{intra}}(t) = \frac{-e}{m} \int_{\Omega} \frac{dr}{\Omega} \sum_{i'i} |\mathrm{C}_{i'i}(t)|^2 Re\, {\varphi_{i'}}^* \times \left(-i\hbar\nabla + \frac{e}{c} A(t)\right) \varphi_{i'} + \delta J_{\mathrm{intra}}(t) \tag{6}$$

where $\delta J_{\mathrm{intra}}(t)$ is the intraband contribution from the pseudopotential. The interband current is calculated as

$$J_{\mathrm{inter}}(t) = J(t) - J_{\mathrm{intra}}(t)\,. \tag{7}$$

We consider inversion asymmetric AB-stacked bilayer T'-$WTe_2$. Figure 1(a) illustrates the crystal structure. We use a rectangular unit cell with the experimental lattice constant [36,37] of $a = 3.483,$ and $b = 6.265$ Å, and the interlayer distance is set as 3.4 Å. A slab approximation is used for the z axis with a distance of 15 Å between the atomic bilayers in the z-direction. For electron-ion interaction, we employ the norm-conserving pseudopotential [38] for W and Te atoms. The adiabatic local density approximation with Perdew-Zunger functional [39] is used for the exchange and correlation. The spin-orbit coupling (SOC) is implemented in a noncollinear mode [40,41]. We use the vector potential of the following waveform,

$$A^{(i)}(t) = -\frac{cE_{\max}}{\omega} f(t)\cos\left\{\omega\left(t - \frac{T_P}{2}\right)\right\} \tag{8}$$

where $\omega$ is the average frequency, $E_{max}$ is the maximum amplitude of the electric field and $T_P$ is the pulse duration. We use the linearly polarized laser field along the x-direction with a time profile of $\cos^4$ envelope, the $f(t)$ is pulse envelope function given as

$$f(t) = \begin{cases} \cos^4\left(\pi \frac{t-T_P/2}{T_P}\right) & 0 \leq t \leq T_P \\ 0 & \text{otherwise} \end{cases} \tag{9}$$

The pulse length is set to 6 optical cycles, and to see the delayed response, the total computation time is set to ~1.5 times the pulse length. In our calculations, we optimized spatial grid sizes and k-points to ensure convergence. The converged real space grid spacing for the $x$ and $y$ directions is $\sim$ 0.20 Å. Since electronic excitations predominantly occur around the pair of Dirac like points in the *k*-space, we employ a non-uniform *k*-mesh which is denser in those regions and coarser sampling in others. For details, refer to the Supplementary Fig. S1. This approach improves accuracy without significantly increasing computational costs. Specifically, we use a non-uniform optimized k-mesh consisting of 523 *k*-points while the time evolution is simulated with a time step of $8.27\times10^{-4}$ fs.

## III. RESULTS AND DISCUSSION

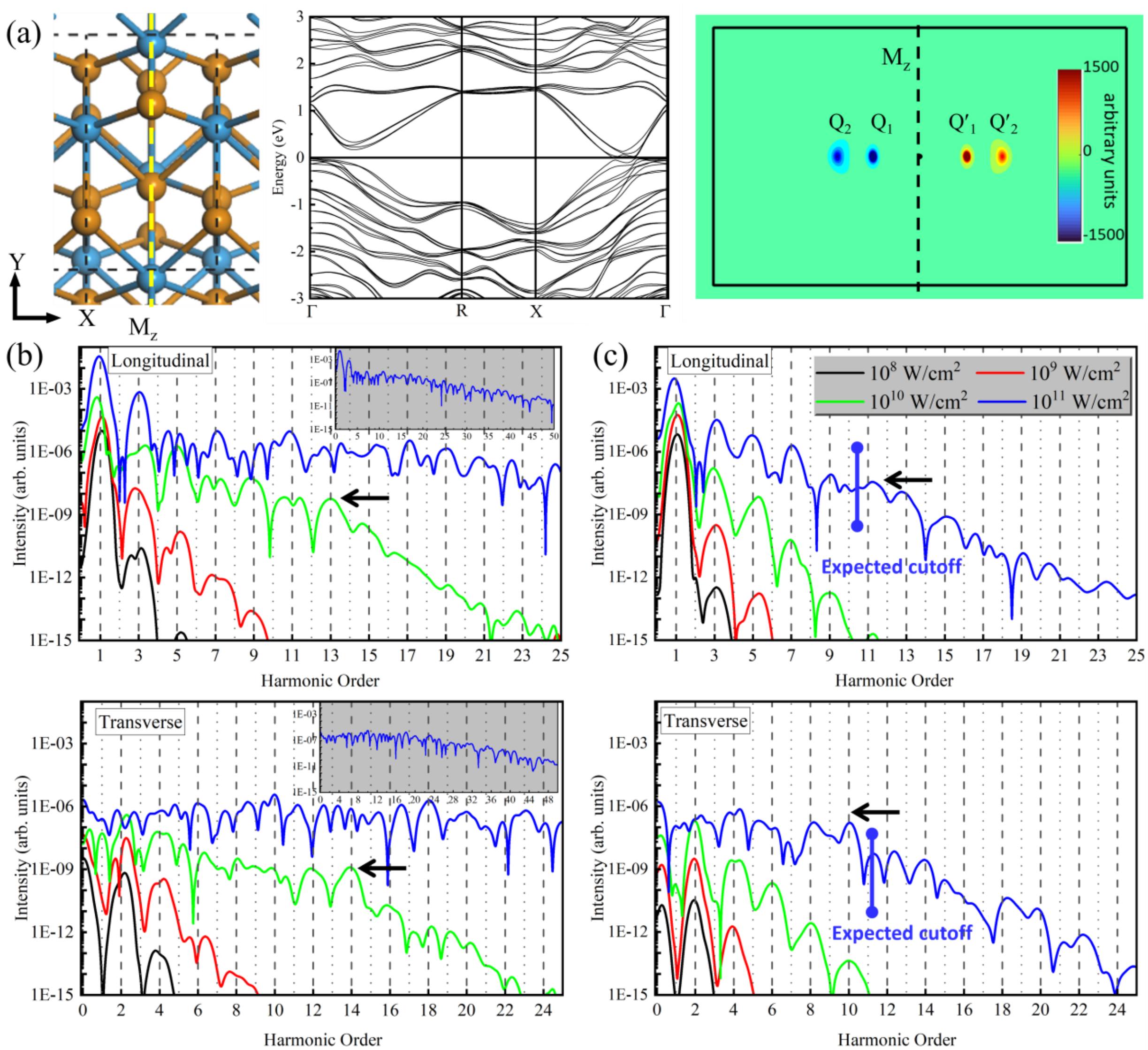

Fig. 1: The details of the crystal symmetry and higher order response of bilayer $WTe_2$, (a) from left to right, crystal structure with mirror plane $M_x$, the band structure including SOC and the Berry-curvature (calculated by OpenMX [42]) indicating the peaks related to the Dirac fermion points. High harmonic spectra at (b) $\hbar\omega$=0.2 eV and (c) $\hbar\omega$=0.4 eV. We multiply $J(t)$ with the envelope function of $\cos^4$ to clarify the harmonics order by suppressing the remaining current after the pulse. The inset in (b) shows the harmonic spectra in an expanded frequency range at $10^{11}$ W/cm$^2$. The arrows in the panels indicate the cutoff. The cutoff in a harmonic spectrum is judged according to the position where the harmonic efficiency decreases after a plateau.

Figure 1(a) presents the crystal and electronic band structure of bilayer $WTe_2$, highlighting two sets of mirror symmetric Dirac-like points associated with large Berry curvature labeled as $Q_{1(2)}$ and $Q'_{1(2)}$. Symmetry analysis reveals that the combination of mirror ($M_x$) and time reversal symmetry (T) effectively acts as space-time-inversion symmetric along the $x$-direction. However, under broken inversion symmetry along the $k_x$ direction, the Berry curvature becomes antisymmetric in momentum space, acting as a Berry curvature dipole [43]. As a result of the interplay of symmetry and Berry curvature [44], the bilayer $WTe_2$ exhibits an even-order transverse response solely under a light polarized along the Berry curvature dipole [8], which is the focus of this study. Multiple higher-order responses can be explicitly analyzed in the frequency space. Therefore, we conduct a comprehensive analysis of the HHG spectra calculated as the modulus square $|F[J(t)]|^2$ of the Fourier transform of the current $J(t)$. Figure 1(b) illustrates the HHG spectra for longitudinal and Hall response at $\hbar\omega$= 0.2 eV. A clear distinction between odd-order (ω, 3ω, 5ω, …) and even-order harmonics (0ω, 2ω, 4ω, …) for the longitudinal and transverse (Hall) spectra, is observed, respectively. Berry curvature plays a dominant role in determining the direction and magnitude of the transverse response. Other effects like band curvature and the Berry connection tend to cancel out due to the symmetries of band dispersion in the transverse direction [45,46]. At intensities ranging from $10^8$ to $10^9$ W/cm$^2$, the HHG spectrum exhibits a typical perturbative behavior characterized by a monotonous decrease in efficiency with increasing harmonic order. With the increase of intensity to $10^{10}$ W/cm$^2$, HHG spectra exhibit a plateau -- a decisive characteristic of a nonperturbative regime -- with a cutoff around 13$^{th}$ and 14$^{th}$ order (~ 2.6 eV) for odd and even responses, respectively. The energy cutoff of harmonic yield has also been discussed in previous model-based Weyl semimetal study [47]. At an intensity of $10^{11}$ W/cm$^2$, the HHG spectra show high energy harmonics without any clear plateau for both longitudinal and Hall harmonics. Such HHG can be considered analogous to plasma harmonics in the gas phase under strong field conditions [48,49]. In the case of $\hbar\omega$= 0.4 eV (Fig. 1(c)), we find a similar trend, while the response appears to be perturbative even at $10^{10}$ W/cm$^2$. On the other hand, harmonic generation shows nonperturbative behavior with the cutoff around ~ 11$^{th}$ order. In both cases of applied frequency, as the intensity increases, the longitudinal and Hall (transverse) harmonic spectra undergo simultaneous transitions from perturbative to nonperturbative regimes and have similar cutoff energies.

The relationship between the cutoff energy and the driving laser wavelength remains a topic of active debate. While some studies suggest a wavelength-independent cutoff position [23,50,51], others propose a linear dependence [52–54]. Our study supports the latter. Since the relation between vector potential amplitude $A_0$ and electric field amplitude $E_0$ is $A_0 \sim \frac{E_0}{\omega}$, where $\omega$ denotes the driving frequency, If we assume the linear relationship, the $A_0$ for the 0.4 eV case at $10^{11}$ W/cm$^2$ is expected to be $(\sqrt{10}/2)$ times to that of the 0.2 eV at $10^{10}$ W/cm$^2$, i.e., the energy cutoff should be ~ 4.1 eV. The linear relationship is evident in the HHG spectra, where the cutoff in Fig. 1(c) (blue lines) is at 4.4 eV, closely aligning with the above-mentioned estimation.

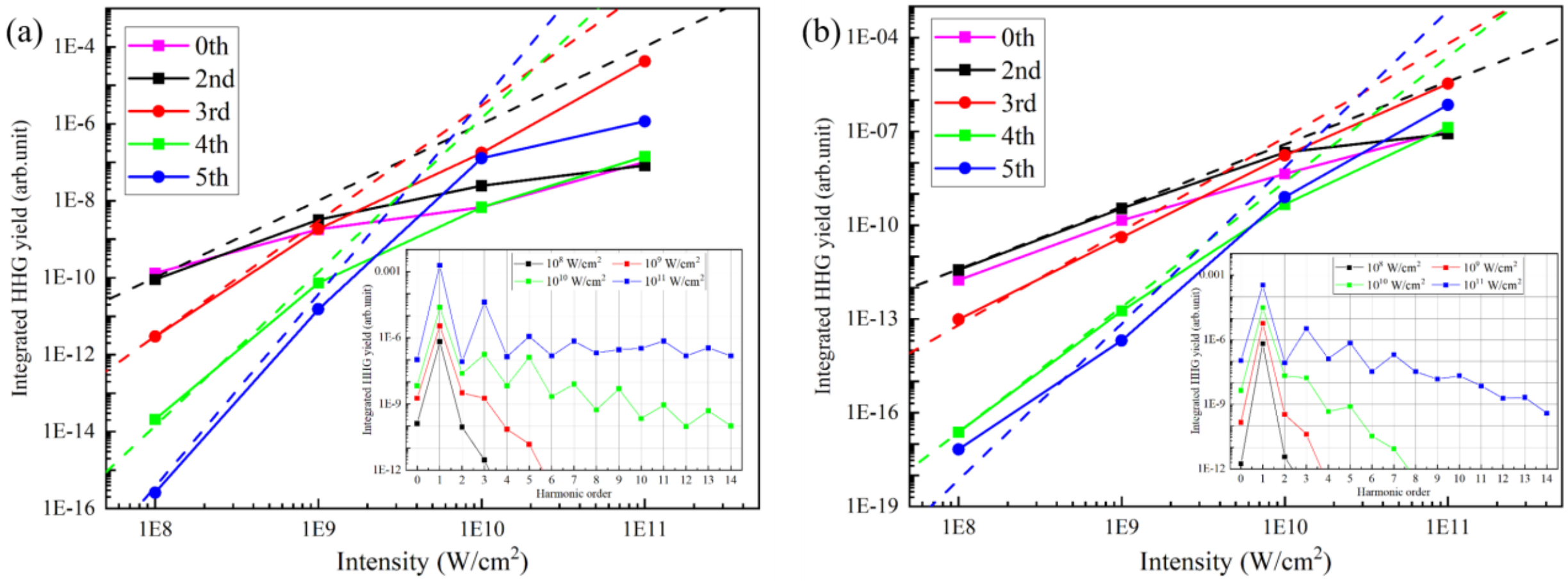

Fig. 2: Logarithmic plot of the integrated harmonic yield of the $0^{th}$, 2nd–5th as a function of the intensity, (a) $\hbar\omega = 0.2$ eV (b) $\hbar\omega = 0.4$ eV. The dashed lines indicate the power law fit, and solid lines with symbols show the calculated results. The inset shows the yields of harmonic order at all intensities. The integrated HHG yield is calculated by integrating the spectral power in a window of two harmonic orders around each harmonic peak since there are only odd and even harmonics in longitudinal and transverse directions, respectively.

Figure 2 shows the integrated harmonic yields as a function of intensity for the longitudinal and Hall responses. Within the perturbative regime, the intensity scaling for the $n$th harmonic should closely follow the power-law $I_L^n$ [55]. At $\hbar\omega = 0.2$ eV within the perturbation regime for intensity $10^8$ - $10^9$ W/cm$^2$, the slight deviation from the expected power law potentially indicates nonlinearity associated with saturable absorption [56]. In the nonperturbative regime, an intensity interval of $10^9$ to $10^{10}$ W/cm$^2$ shows further deviation, and beyond $10^{10}$ W/cm$^2$ we find a complete deviation from the power law, which is consistent with the appearance of the plateau [Fig. 1(b) and the inset of Fig. 2(a)]. With $\hbar\omega = 0.4$ eV (Fig. 2(b)), the deviation from the power law is less prominent than in the 0.2 eV case, which coincides with the slower appearance of the plateau [Fig. 1(c) and the inset of Fig. 2(b)], indicating that the perturbative behavior persists to higher intensities for higher frequencies. Similar power-law divergences have been reported in previous HHG studies on semimetals and semiconductors under intense fields [17,49,56,57]. Additionally, within the perturbation regime, the transverse even-order response induced by the Berry curvature exhibits a yield comparable to the corresponding odd-order longitudinal response. However, in the nonperturbative regime, the even-order Hall harmonics are systematically weaker than the neighboring odd-order longitudinal harmonics. These results contrast with a previous report [17], where comparable even and odd-order responses have been reported, and its origin has been attributed to intraband motions and the Berry curvature. Figure 2 also shows the zero-order component, which is part of the second-order optical response and, therefore, follows the same symmetry constraints. Notably, the zero-order component deviates from power law behavior as observed in the second-order harmonic. Within the framework of perturbation theory, the zero-order component can arise from mechanisms such as shift currents and optical rectification, with shift currents likely dominating for frequencies above the band gap [58,59]. Although the shift current is not the primary focus of this work, the presence of a net direct current often referred to as the linear photogalvanic effect (LPGE), holds significant promise for applications in photovoltaic and optoelectronic devices. It is important to note that the LPGE is fundamentally distinct from the injection current associated with the circular photogalvanic effect (CPGE) by the absorption of circularly polarized photons [60,61].

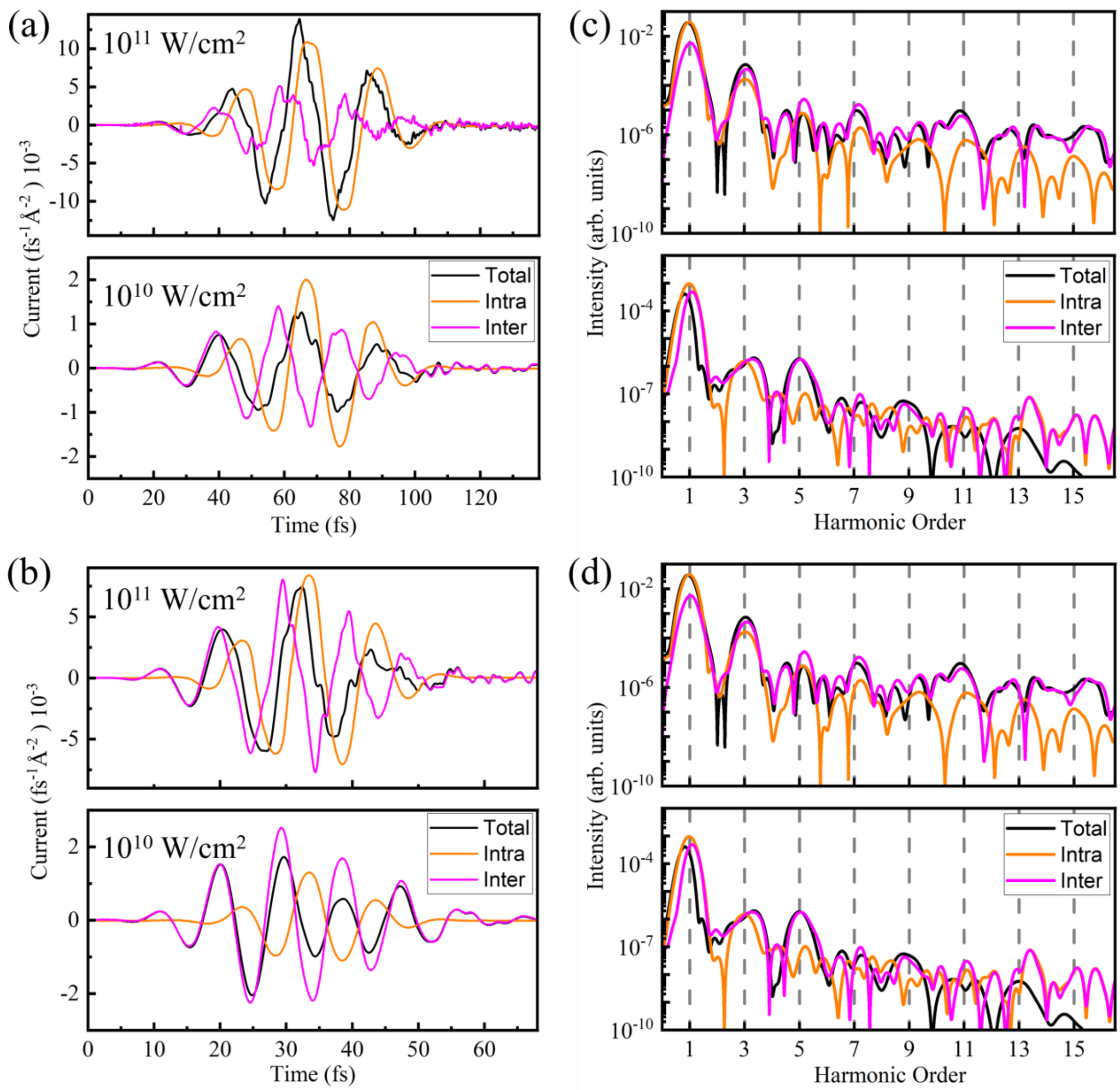

Fig. 3: Contribution of inter and intraband current to longitudinal current (a) 0.2 eV and (b) 0.4 eV. Decompositions of the corresponding HHG into inter/intraband (c) 0.2 eV and (d) 0.4 eV.

Let us now examine the inter- and intraband contributions to elucidate the origin of longitudinal and Hall responses in the present case and the difference from the previous study [17], since the nonperturbative higher-order response in solids arises in general from the combination of intraband and interband contributions to the nonlinear current [50,62–66]. Band dispersion influences intraband contributions, while interband contribution is due to induced polarization between valence and conduction bands. To separate the two contributions, we project time-dependent Kohn-Sham orbitals onto instantaneous eigenstates of the time-dependent Hamiltonian, as detailed in ref. [67]. Figure 3 (a) and (b) compare the inter- and intraband contributions to longitudinal currents for 0.2 eV and 0.4 eV, respectively. Our results show that the intra-band contribution to longitudinal current is minimal at weak intensities ($I \leq 10^9$ W/cm$^2$, not shown). However, the strength of the intra-band current becomes comparable to the interband with increasing intensity ($I \geq 10^{10}$ W/cm$^2$), indicating that an intensity-dependent interplay between the inter- and intraband contributions forms harmonic spectra. This observation is in contrast to previous reports for graphene under strong THz or mid-infrared fields [49,68], where the intraband current is dominant. Due to the larger vector potential (at the same intensity), the intraband contributions are more significant at $\hbar\omega$=0.2 eV. The phase of intraband current does not change across the varying intensities. On the other hand, a notable variation in the phase of interband currents with the driving intensity is observed, suggesting a significant change in resonant excitation mechanisms. This evolving phase difference between inter/intraband currents leads to a complex interplay between interband polarization and nonlinear motion in the energy bands, giving rise to the net total current.

Figures 3(c, d) show the intra- and interband HHG spectra, obtained as the modulus squared of the Fourier transform of the intra- and interband currents, respectively. The intraband contribution experiences a decline after the first order, with the interband polarization taking precedence. However, near the plateau region the two contributions become comparable, indicating that the plateau is formed due to interference between inter and intra harmonics in both 0.2 and 0.4 eV cases. The interband transition mainly arises through multiphoton resonance. However, the intraband motions, followed by the inter-band excitations, can further excite the electron in the conduction band. Such excitation in turn can generate high-energy harmonic spectra through electron-hole recombinations. Such mechanisms are particularly stronger with the highly dispersive bands where a small change in momentum ($Q + A(t)$) can induce larger excitations and high energy harmonics. The presence of substantial intraband contribution at the first order and near the plateau region may arise from two reasons. Firstly, according to the Bloch acceleration theorem, the $\hbar\omega$ drives intraband motion, and carriers traverse in reciprocal space, along the $Q + A(t)$ trajectory from the Dirac-like point. As a result, their velocities rapidly switch between approximately $\pm vf$ [55,69,70], contributing significantly to the intraband harmonic spectrum. Secondly, the anharmonic dispersion of Dirac-like bands can also lead to a significant intraband contribution [71]. At a strong intensity of $10^{11}$ W/cm$^2$ for $\hbar\omega = 0.2$ eV, the HHG spectra exhibit distinct characteristics, displaying a noisy spectrum with strong spectral contributions across all energies. This can be attributed to the presence of numerous electron-hole recombination channels [23].

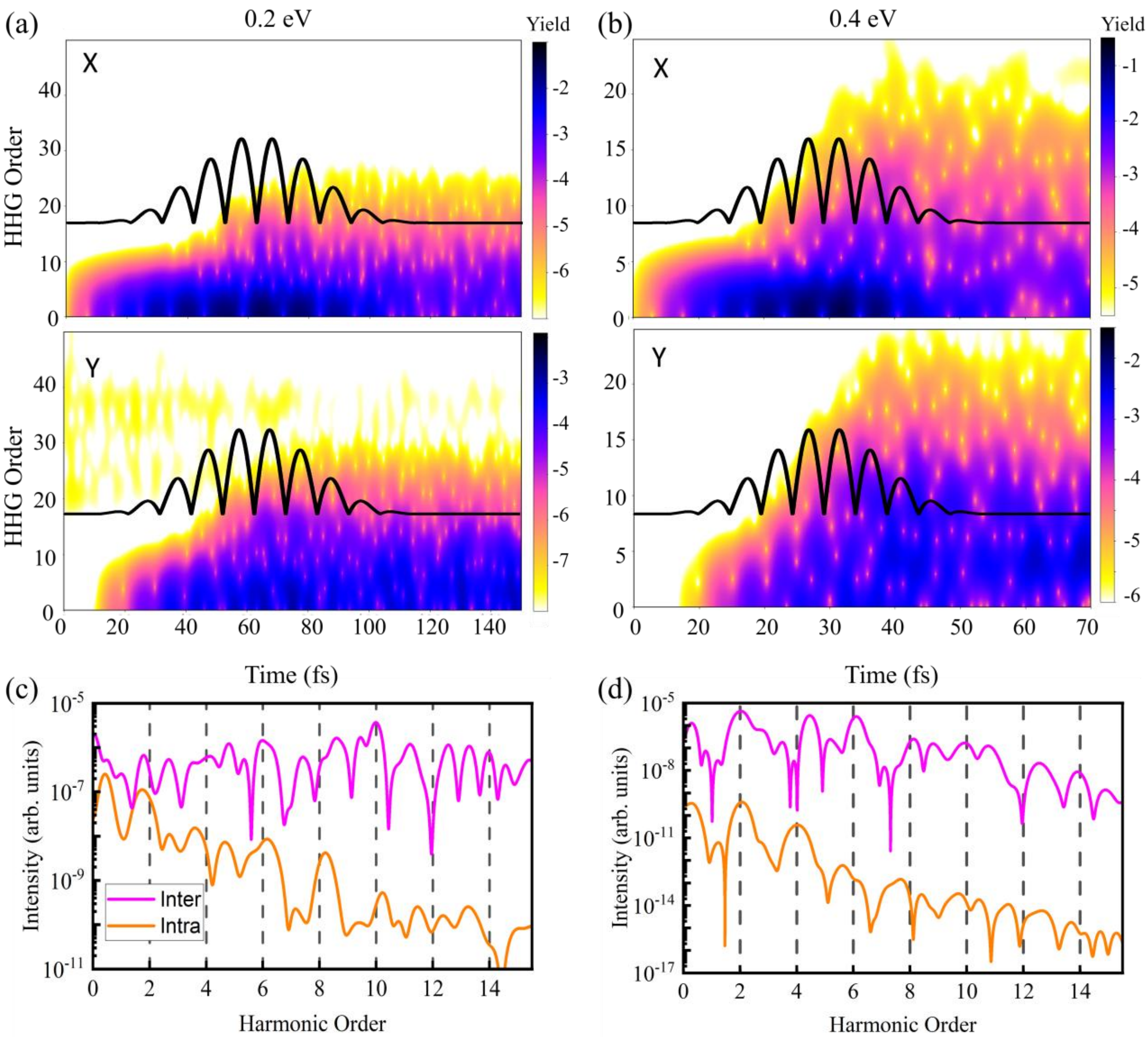

Fig. 4 Time-frequency analysis of the HHG process (a) 0.2 eV-$10^{10}$ W/cm$^2$ and (b) 0.4 eV-$10^{11}$ W/cm$^2$. We used a window of 50 fs to perform the Gabor transformation. The black curve is the profile of the applied vector potential $A(t)$ of the laser pulse. Decompositions of harmonic spectra into inter/intraband contributions (c) 0.2 eV and (d) 0.4 eV for anomalous Hall harmonics at maximum intensity of $10^{11}$ W/cm$^2$.

Figures 4(a, b) present the spectrograms for the HHG shown in Fig. 1, specifically for the cases where the HHG plateau and energy cutoff are observed. It shows a subtle temporal shift between longitudinal and Hall harmonic spectra. The longitudinal harmonics (upper panels) in Fig. 4(a), up to the 13th order are primarily generated during the pulse, suggesting a significant intraband contribution. This behavior is consistent with the inter/intraband current contributions shown in Fig. 3. On the other hand, in the transverse direction, all harmonic emissions show a time delay, indicating the generation and recombination of electrons and holes. In addition, the harmonic emission persists beyond the pulse duration, demonstrating that the Hall harmonics entirely arise from interband polarization. Fig. 4(b) depicts similar behavior for longitudinal harmonics at 0.4 eV with an intensity of I=$10^{11}$ W/cm$^2$. However, a slight variation can be observed in the Hall HHG spectrogram, suggesting a frequency dependence for Hall harmonics. The calculated results show the absence of the intraband contributions to Hall harmonics. Therefore, the origin of the Hall harmonics can be associated with the large interband Berry curvature of the Dirac-like semimetal [29]. Figures 4 (c, d) display inter and intraband contributions to Hall harmonics at 0.2 and 0.4 eV, with a maximum intensity of $10^{11}$ W/cm$^2$. Contrary to the longitudinal response, the intraband contributions are smaller than the interband ones by orders of magnitude even in the nonperturbative regime, which indicates that the Hall harmonics are almost entirely due to the interband contributions.

Previous work on the monolayer of $MoS_2$ [16] and Weyl semimetal β-$WP_2$ [17] has reported that the perpendicular even harmonics are associated with intraband current in the presence of the Berry curvature. These studies used semiclassical equations of motion within the independent particle approximation, incorporating the Berry curvature term to explain the anomalous Hall current. However, in many electron systems under the adiabatic approximation, Berry curvature of band index n can defined as $\Omega_{\mu\nu}^{n}(\boldsymbol{R}) = i\sum_{n'\neq n}\frac{\langle n|\partial H/\partial R^{\mu}|n'\rangle\langle n'|\partial H/\partial R^{\nu}|n\rangle-(\nu\leftrightarrow\mu)}{(\varepsilon_n-\varepsilon_{n'})^2}$ ($\nu\mu$ represent the cartesian coordinates) which is fundamentally a projection operation arising from residual coupling between bands [72–76]. However, under the dipole approximation, the semiconductor Bloch equation (SBE) can provide analytical insight. By carefully disentangling different current density components, we identify the origin of Berry curvature-induced Hall currents (see Supplementary Section S3). The analytical solution show that Berry curvature term can only be obtained from the expansion of the interband (off-diagonal) current density. Although the total current as an observable is gauge invariant, the decomposition into different contributions is not unique and depends on the chosen framework [75]. Nonetheless, our analysis shows that, regardless of the chosen gauge (length, velocity, or Houston basis), the Berry curvature term (perpendicular to the field) is an inherent part of the interband current which is also true for our method where we define the intraband current by Eq. (6). We also extend the current expression for the case where interband dynamics are negligibly small, effectively reducing the system to a single band. This framework assumes that a small fraction of electrons tunnels from the valence to the conduction band, leaving only one band significantly populated. Under this approximation, only the diagonal component of the decomposed current survives (see Supplementary Section S3 (D)). Notably, this diagonal current correspond to the semiclassical equation of motion for carriers within a single band, where the Berry curvature term is considered as an intraband contribution. Since the Berry curvature term is assigned to interband and intraband contributions in multiband and single-band cases, respectively, it is important to emphasize that this assignment is arbitrary rather than a physically essential distinction. We also conducted a SBE study on our system (Supplementary Section S4) and TDDFT study on $MoS_2$ (see Supplementary Section S5). One can deduce two main conclusions from this discussion: First, it is important to comprehend that single-band models and multiband schemes (such as TDDFT and SBE) offer distinct yet complementary perspectives. While they might at first seem to lead to contradictory conclusions, they are, in fact compatible. Secondly, since the Berry curvature effects are inherently linked to interband coherence, further analysis, including the separation of intra- and interband contributions to longitudinal currents along with the excitation mechanism, will be necessary to get a more complete picture of the system.

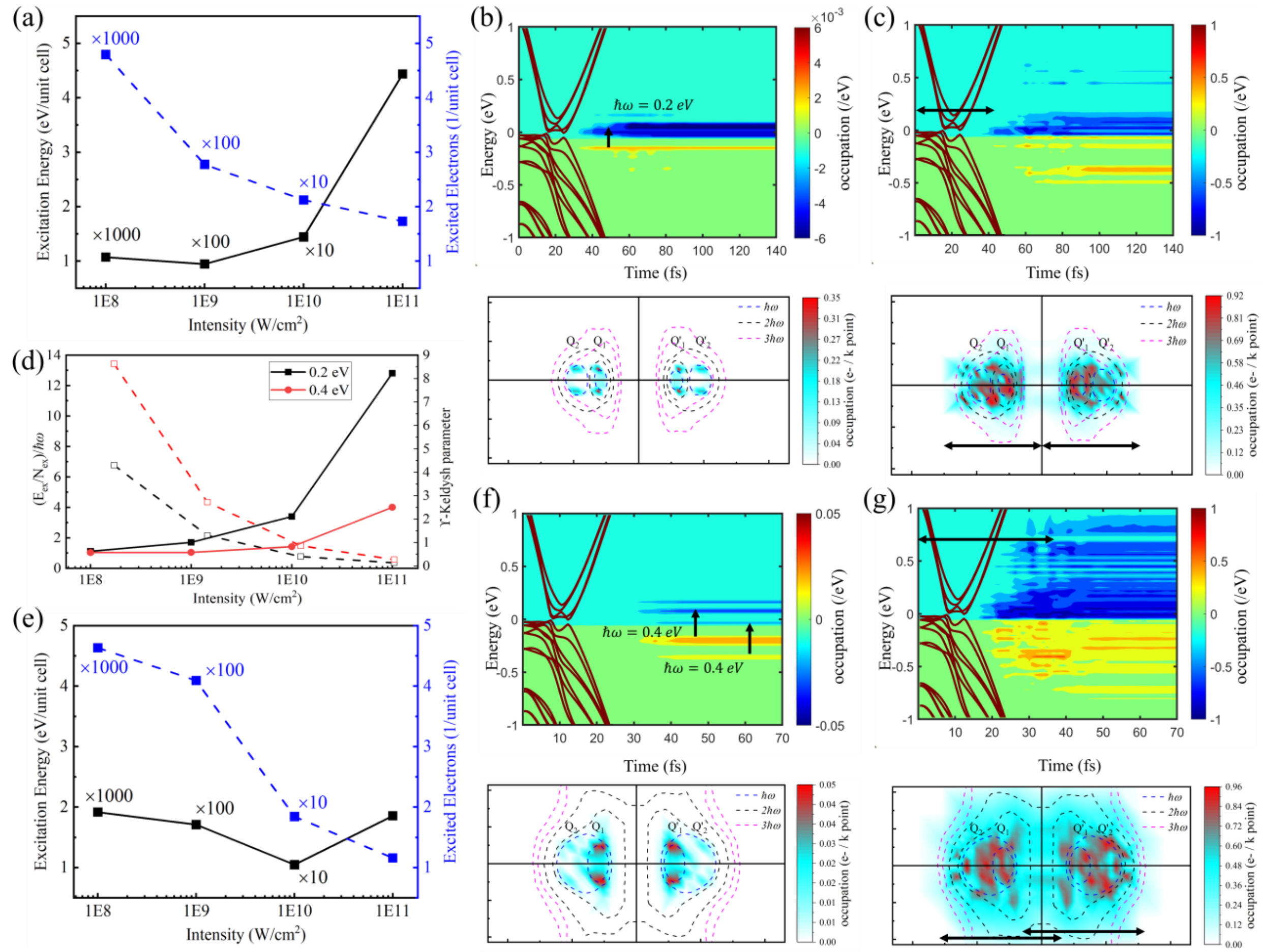

Fig. 5: The change in excitation mechanism with field strength. (a) Excitations at $\hbar\omega$ = 0.2 eV, the solid lines show excitation energy, and dashed lines are for excited electrons. (b) The temporal evolution of charge excitations projected on the initial field-free Kohn-Sham orbitals with band structure at X-Γ line shown by brown solid lines (upper panel), and the distribution of *k*-resolved electron populations of $CB_1$ at the end of the pulse (down panels) for intensity of $10^8$ W/cm$^2$. (c) Same as (b) but for the intensity of $10^{10}$ W/cm$^2$. The black arrows indicate the vector potential amplitude at specific intensities. The dashed curves indicate instances where the energy difference between two bands aligns with the photon energy n$\hbar\omega$ ($\hbar\omega$ = 0.2 eV) where n = 1, 2 and 3. (d) Change in photon excitation energy and Keldysh parameter (γ). The solid and dashed lines represent $(E_{ex}/N_{ex})/\hbar\omega$ and Keldysh parameter respectively. (e)-(g) are for the $\hbar\omega$ = 0.4eV, (e) excitations as described for (a), the temporal evolution of excitations (upper panes) and k-resolved population (down panels) at (f) $10^8$ W/cm$^2$ and (g) $10^{11}$ W/cm$^2$.

The longitudinal and Hall current responses are intricately tied to the dynamics of the excited carriers. Let us now delve into the role of excitation at different field strengths. Figure 5(a) presents the normalized excitation energy and the excited electron density at $\hbar\omega$ = 0.2 eV. The excitation energy is evaluated as the difference between the energy density integrated over the unit cell at the end of the pulse and that in the initial ground state. The excitation energy depends almost linearly on intensity up to $10^{10}$ W/cm$^2$. However, at $10^{11}$ W/cm$^2$, we observe a significant nonlinear increase, indicating a shift in the excitation mechanism (later explained in Fig. 5(d)). The sublinear intensity dependence of excited carriers is attributed to saturable absorption, which also accounts for the slight deviation from the power law between $10^8$ - $10^9$ W/cm$^2$ in Fig. 2. The projected excitation at $10^8$ and $10^{10}$ W/cm$^2$ is depicted in Fig. 5(b) and 5(c), respectively. Single-photon absorption is prominent at $10^8$ W/cm$^2$ [Fig. 5(b)], while it changes to two-photon absorption as the intensity increases to $10^9$ W/cm$^2$ (not shown). As evident from the lower panel of Fig. 5(b), single-electron excitations are predominantly localized around the $Q_{1(2)}/Q'_{1(2)}$ Dirac points where the Berry curvature exhibits sharp peaks (see Fig. 1). This underscores the critical role of the Berry curvature dipole in driving the strong second-order Hall response, which exceeds the third-order response in the perturbative regime. Under more intense fields ($\geq 10^{10}$ W/cm$^2$), low-energy states become saturated through single-photon excitation (< 60 fs in Fig. 5(c)), and subsequently, high-energy states are involved through multiphoton excitation. It is important to consider the energy gain of charge carriers through intra-band motions driven by vector potential. Our calculations indicate that the harmonics resulting from this contribution can reach up to ~ 2.4 eV. As shown in Supplementary Fig. S2, the excitation occupation shows sizable peaks ranging from -1.98 eV in the valence band to ~2.22 eV in the conduction band. Therefore, the cutoff harmonics can be associated with the interband transition within this range. Due to the engagement of intraband motions and multiphoton excitations, excited electrons are no longer confined to $Q_{1(2)}/Q'_{1(2)}$ Dirac points but encompass a wider region of the Brillouin zone (BZ). The hotspot of the population emerges off-resonant, situated between one/multiphoton absorption resonances, which explains the noisy spectrum at strong intensity [Fig. 3(c)]. The excitation energies along with the Keldysh parameter ($\gamma$) are plotted in Fig. 5(d). The $\gamma$ serves as a commonly used metric for determining the excitation regime: $\gamma >> 1$ ($\gamma << 1$) corresponds to the multiphoton (tunneling) regime. It should be noted that the Keldysh theory was originally developed for quasi-static fields compared to the electronic time scale ($\hbar\omega << E_g$), which may not directly apply to our case. However, the $\gamma$ along with the computed values of $(E_{ex}/N_{ex})/\hbar\omega$, can offer valuable insights into the excitation regime. In the case of 0.2 eV, the excitation mechanism undergoes a rapid transition from single/multiphoton to tunneling as intensity increases, which is also corroborated the $\gamma$ values. In contrast, for 0.4 eV, single-photon absorption prevails up to ~ $10^{10}$ W/cm$^2$, and the $\gamma > 0.5$ values also support the multiphoton regime [77]. These results demonstrate that the excitation process under the 0.2 eV field exhibits stronger nonlinearity compared to the 0.4 eV field.

The normalized excitation energy and the excited electrons for $\hbar\omega$=0.4 eV in Fig. 5(e) exhibit similar behavior to that observed in the 0.2 eV case. Figure 5(f) clearly demonstrates multiband single-photon excitation at $10^8$ W/cm$^2$. This occurs due to the equivalence of photon energy (0.4 eV) with the energy gap between different bands, enabling resonant absorption. We note that these excitations manifest themselves in two distinct forms: transitions from the valence band maximum ($VB_1$) to higher energy conduction bands ($CB_2$), and transitions from lower energy valence bands ($VB_{2/3}$) to the conduction band minimum ($CB_1$). The excited population for an intensity of $10^8$ W/cm$^2$at the end of the pulse indicates that interband excitations remain concentrated in the vicinity of the Dirac point, as depicted in Fig. 5(f). The energy gain of charge carriers through intra-band motions can contribute up to ~ 4.0 eV which also matches with the cutoff energy. The analysis of excited carrier occupation can be seen in Supplementary Fig. S2. Additionally, at the maximum intensity of $10^{11}$ W/cm$^2$, the off-resonant excitation spreads across the BZ (see Fig. 5(g)), indicating substantial intraband motion.

In the end, it is important to address a few key points. Berry curvature plays a critical role in the observed high-order Hall harmonic responses under broken inversion symmetry. Specifically, the second-order harmonic can be linked to the Berry curvature dipole, and higher-order contributions can be connected to Berry curvature multipoles within perturbative limits [78]. However, the nonperturbative responses under strong fields, as described in our study, where high-order Hall harmonics feature plateaus and cutoffs, go beyond the scope of the perturbative theory of Berry curvature dipole. Since our study highlights the importance of interband coherences, and higher energy bands involvement far from Dirac points necessitate the use of advanced techniques like first-principles approaches (e.g., TDSE with TDDFT). Simplified approaches, such as tight-binding models using minimal-band frameworks, cannot fully capture the complex dynamics of the nonperturbative regime. Thus, extending single-band (one-electron) models to multiband (multi-electron) frameworks is essential for identifying interband and intraband contributions and their roles in harmonic generation.

**IV. CONCLUSION**

The strong field optical response of the time-reversal pair of coupled Dirac fermions in bilayer $WTe_2$ is explored by using TDDFT. The calculated results show changes in carrier dynamics with varying applied field strength, unveiling intriguing dynamics in two distinct regimes. In the weaker field regime, the even-order Hall harmonics induced by the Berry curvature show yields comparable to corresponding odd-order longitudinal responses. The inter-band resonant excitations dominate the longitudinal high harmonic responses. The excitation mechanism changes from single-photon absorption at low intensities to multiphoton absorption as the available states near the Dirac-like points becomes saturated. The intraband contributions to the longitudinal currents increase with intensity, highlighting an intensity-dependent interplay between interband and intraband contributions. At higher frequencies, the effect of multiband excitation becomes prominent due to the availability of electrons for resonant excitation. However, the nonlinear behaviors, including tunneling and multiphoton excitations, slows with increasing frequency. Interestingly, when adopting an all-band perspective enabled by TDDFT, the anomalous Hall responses are found to be almost entirely attributed to the interband processes in both perturbative and nonperturbative regimes, highlighting the importance of frequency dependence and interband transitions in nonlinear Hall responses. Overall, the results show the importance of interband mechanisms for both longitudinal and Hall HHG. Our comprehensive investigation paves the way for a deeper understanding of nonlinear phenomena in Weyl-like materials.

**Supplementary Information.**

The supplementary information details the $k$-point sampling strategy for analyzing excitations in the Dirac region. It includes analyses of laser-induced excitation occupations, Semiconductor Bloch Equations calculations for current decomposition, analytical discussions, and TDDFT calculations of Hall harmonics for the MoS2 monolayer explaining the Berry curvature contributions. Each section describes the methods, materials, and parameters used.

**Notes**

The authors declare no competing financial interest.

**ACKNOWLEDGMENT**

The work in China was supported by the National Key R&D program of China (Grant Nos. 2022YFA1402903 and 2024YFA1409101), the National Natural Science Foundation of China (Grant No. 12374059). Computational time was supported by the Center for Computational Science and Engineering of Southern University of Science and Technology, and the Major Science and Technology Infrastructure Project of Material Genome Big-science Facilities Platform supported by Municipal Development and Reform Commission of Shenzhen. Part of this work was supported by the Quantum Science Center of Guangdong-Hong Kong-Macao Greater Bay Area (Guangdong). This research in Japan is supported by JSPS KAKENHI Grant No. JP20H05670. This research is also partially supported by MEXT Quantum Leap Flagship Program (MEXT Q-LEAP) under Grant No. JPMXS0118068681 and JPMXS0118067246, JSPS KAKENHI Grant Nos. JP20H02649, JP22K13991, JP22K18982, JP24H00427, JP24K01224, 24K23024, and 25K17367 and JST-CREST under Grant No. JP-MJCR16N5. The numerical calculations are carried out using the computer facilities of the Fugaku through the HPCI System Research Project (Project ID: hp220120, hp240124), SGI8600 at Japan Atomic Energy Agency (JAEA), and Wisteria at the University of Tokyo under Multidisciplinary Cooperative Research Program in CCS, University of Tsukuba and JSPS KAKENHI Grant, JP22K13991.

## REFERENCES

[1] J.-X. Zhang, Z.-Y. Wang, and W. Chen, Disorder-induced anomalous Hall effect in type-I Weyl metals: Connection between the Kubo-Streda formula in the spin and chiral basis, Phys. Rev. B **107**, 125106 (2023).

[2] Y. Tokura, K. Yasuda, and A. Tsukazaki, Magnetic topological insulators, Nat. Rev. Phys. **1**, 2 (2019).

[3] L. L. Tao and E. Y. Tsymbal, Persistent spin texture enforced by symmetry, Nat. Commun. **9**, 1 (2018).

[4] M. U. Farooq, L. Xian, and L. Huang, Spin Hall effect in two-dimensional InSe: Interplay between Rashba and Dresselhaus spin-orbit couplings, Phys. Rev. B **105**, 245405 (2022).

[5] M. U. Farooq, Z. Gui, and L. Huang, Spontaneous spin momentum locking and anomalous Hall effect in ${\mathrm{BiFeO}}_{3}$, Phys. Rev. B **107**, 075202 (2023).

[6] C.-Z. Chang, C.-X. Liu, and A. H. MacDonald, *Colloquium* : Quantum anomalous Hall effect, Rev. Mod. Phys. **95**, 011002 (2023).

[7] N. Nagaosa, J. Sinova, S. Onoda, A. H. MacDonald, and N. P. Ong, Anomalous Hall effect, Rev. Mod. Phys. **82**, 1539 (2010).

[8] Q. Ma et al., Observation of the nonlinear Hall effect under time-reversal-symmetric conditions, Nature **565**, 7739 (2019).

[9] I. Sodemann and L. Fu, Quantum Nonlinear Hall Effect Induced by Berry Curvature Dipole in Time-Reversal Invariant Materials, Phys. Rev. Lett. **115**, 216806 (2015).

[10] P. He, H. Isobe, D. Zhu, C.-H. Hsu, L. Fu, and H. Yang, Quantum frequency doubling in the topological insulator Bi2Se3, Nat. Commun. **12**, 1 (2021).

[11] D. Kumar, C.-H. Hsu, R. Sharma, T.-R. Chang, P. Yu, J. Wang, G. Eda, G. Liang, and H. Yang, Room-temperature nonlinear Hall effect and wireless radiofrequency rectification in Weyl semimetal TaIrTe4, Nat. Nanotechnol. **16**, 4 (2021).

[12] Z. Z. Du, H.-Z. Lu, and X. C. Xie, Nonlinear Hall effects, Nat. Rev. Phys. **3**, 11 (2021).

[13] H. Wang and X. Qian, Ferroelectric nonlinear anomalous Hall effect in few-layer WTe2, Npj Comput. Mater. **5**, 1 (2019).

[14] C.-P. Zhang, X.-J. Gao, Y.-M. Xie, H. C. Po, and K. T. Law, Higher-order nonlinear anomalous Hall effects induced by Berry curvature multipoles, Phys. Rev. B **107**, 115142 (2023).
[15] S. Roy and A. Narayan, Non-linear Hall effect in multi-Weyl semimetals, J. Phys. Condens. Matter **34**, 385301 (2022).
[16] H. Liu, Y. Li, Y. S. You, S. Ghimire, T. F. Heinz, and D. A. Reis, High-harmonic generation from an atomically thin semiconductor, Nat. Phys. **13**, 3 (2017).
[17] Y.-Y. Lv et al., High-harmonic generation in Weyl semimetal β-WP2 crystals, Nat. Commun. **12**, 6437 (2021).
[18] A. J. Uzan-Narovlansky et al., Observation of interband Berry phase in laser-driven crystals, Nature **626**, 66 (2024).
[19] A. Bharti, M. S. Mrudul, and G. Dixit, High-harmonic spectroscopy of light-driven nonlinear anisotropic anomalous Hall effect in a Weyl semimetal, Phys. Rev. B **105**, 155140 (2022).
[20] L. Medic, J. Mravlje, A. Ramšak, and T. Rejec, High-harmonic generation in semi-Dirac and Weyl semimetals with broken time-reversal symmetry: Exploration of the merging of Weyl nodes, Phys. Rev. B **109**, 205130 (2024).
[21] H. K. Avetissian, V. N. Avetisyan, B. R. Avchyan, and G. F. Mkrtchian, High-order harmonic generation in three-dimensional Weyl semimetals with broken time-reversal symmetry, Phys. Rev. A **106**, 033107 (2022).
[22] A. Bharti and G. Dixit, Non-perturbative nonlinear optical responses in Weyl semimetals, Appl. Phys. Lett. **125**, 051104 (2024).
[23] N. Tancogne-Dejean, O. D. Mücke, F. X. Kärtner, and A. Rubio, Impact of the Electronic Band Structure in High-Harmonic Generation Spectra of Solids, Phys. Rev. Lett. **118**, 087403 (2017).
[24] H. Xia et al., Nonlinear optical signatures of topological Dirac fermion, Sci. Adv. **10**, eadp0575 (2024).
[25] T. Higuchi, C. Heide, K. Ullmann, H. B. Weber, and P. Hommelhoff, Light-field-driven currents in graphene, Nature **550**, 7675 (2017).
[26] S. N. Shevchenko, S. Ashhab, and F. Nori, Landau–Zener–Stückelberg interferometry, Phys. Rep. **492**, 1 (2010).
[27] M. Baudisch et al., Ultrafast nonlinear optical response of Dirac fermions in graphene, Nat. Commun. **9**, 1 (2018).
[28] A. Tiwari et al., Giant c-axis nonlinear anomalous Hall effect in Td-MoTe2 and WTe2, Nat. Commun. **12**, 1 (2021).
[29] K. Kang, T. Li, E. Sohn, J. Shan, and K. F. Mak, Nonlinear anomalous Hall effect in few-layer WTe2, Nat. Mater. **18**, 4 (2019).
[30] G. F. Bertsch, J.-I. Iwata, A. Rubio, and K. Yabana, Real-space, real-time method for the dielectric function, Phys. Rev. B **62**, 7998 (2000).
[31] *Salmon Official Website*, http://salmon-tddft.jp.
[32] M. Noda et al., SALMON: Scalable Ab-initio Light–Matter simulator for Optics and Nanoscience, Comput. Phys. Commun. **235**, 356 (2019).
[33] T. Otobe, M. Yamagiwa, J.-I. Iwata, K. Yabana, T. Nakatsukasa, and G. F. Bertsch, First-principles electron dynamics simulation for optical breakdown of dielectrics under an intense laser field, Phys. Rev. B **77**, 165104 (2008).
[34] C. J. Pickard and F. Mauri, Nonlocal Pseudopotentials and Magnetic Fields, Phys. Rev. Lett. **91**, 196401 (2003).

[35] S. Ismail-Beigi, E. K. Chang, and S. G. Louie, Coupling of Nonlocal Potentials to Electromagnetic Fields, Phys. Rev. Lett. **87**, 087402 (2001).
[36] M. N. Ali et al., Large, non-saturating magnetoresistance in WTe2, Nature **514**, 7521 (2014).
[37] C. H. Naylor et al., Large-area synthesis of high-quality monolayer 1T'-WTe2 flakes, 2D Mater. **4**, 021008 (2017).
[38] I. Morrison, D. M. Bylander, and L. Kleinman, Nonlocal Hermitian norm-conserving Vanderbilt pseudopotential, Phys. Rev. B **47**, 6728 (1993).
[39] J. P. Perdew and Y. Wang, Accurate and simple analytic representationof the electron-gas correlation energy, Phys. Rev. B **45**, 13244 (1992).
[40] U. von Barth and L. Hedin, A local exchange-correlation potential for the spin polarized case. i, J. Phys. C Solid State Phys. **5**, 1629 (1972).
[41] T. Oda, A. Pasquarello, and R. Car, Fully Unconstrained Approach to Noncollinear Magnetism: Application to Small Fe Clusters, Phys. Rev. Lett. **80**, 3622 (1998).
[42] T. Ozaki, Variationally optimized atomic orbitals for large-scale electronic structures, Phys. Rev. B **67**, 155108 (2003).
[43] J.-S. You, S. Fang, S.-Y. Xu, E. Kaxiras, and T. Low, Berry curvature dipole current in the transition metal dichalcogenides family, Phys. Rev. B **98**, 121109 (2018).
[44] T. T. Luu and H. J. Wörner, Measurement of the Berry curvature of solids using high-harmonic spectroscopy, Nat. Commun. **9**, 916 (2018).
[45] K. Kaneshima, Y. Shinohara, K. Takeuchi, N. Ishii, K. Imasaka, T. Kaji, S. Ashihara, K. L. Ishikawa, and J. Itatani, Polarization-Resolved Study of High Harmonics from Bulk Semiconductors, Phys. Rev. Lett. **120**, 243903 (2018).
[46] S. Lai et al., Third-order nonlinear Hall effect induced by the Berry-connection polarizability tensor, Nat. Nanotechnol. **16**, 869 (2021).
[47] A. Bharti and G. Dixit, Role of topological charges in the nonlinear optical response from Weyl semimetals, Phys. Rev. B **107**, 224308 (2023).
[48] F. Brunel, Harmonic generation due to plasma effects in a gas undergoing multiphoton ionization in the high-intensity limit, JOSA B **7**, 521 (1990).
[49] M. Taucer et al., Nonperturbative harmonic generation in graphene from intense midinfrared pulsed light, Phys. Rev. B **96**, 195420 (2017).
[50] S. Ghimire, A. D. DiChiara, E. Sistrunk, P. Agostini, L. F. DiMauro, and D. A. Reis, Observation of high-order harmonic generation in a bulk crystal, Nat. Phys. **7**, 138 (2011).
[51] T. T. Luu, M. Garg, S. Y. Kruchinin, A. Moulet, M. T. Hassan, and E. Goulielmakis, Extreme ultraviolet high-harmonic spectroscopy of solids, Nature **521**, 498 (2015).
[52] M. Hohenleutner, F. Langer, O. Schubert, M. Knorr, U. Huttner, S. W. Koch, M. Kira, and R. Huber, Real-time observation of interfering crystal electrons in high-harmonic generation, Nature **523**, 572 (2015).
[53] C. Yu, S. Jiang, and R. Lu, High order harmonic generation in solids: a review on recent numerical methods, Adv. Phys. X **4**, 1562982 (2019).
[54] A. S. Emelina, M. Y. Emelin, and M. Y. Ryabikin, Wavelength scaling laws for high-order harmonic yield from atoms driven by mid- and long-wave infrared laser fields, JOSA B **36**, 3236 (2019).
[55] O. Schubert et al., Sub-cycle control of terahertz high-harmonic generation by dynamical Bloch oscillations, Nat. Photonics **8**, 2 (2014).
[56] N. Yoshikawa, T. Tamaya, and K. Tanaka, High-harmonic generation in graphene enhanced by elliptically polarized light excitation, Science **356**, 736 (2017).

[57] H. Liu, Y. Li, Y. S. You, S. Ghimire, T. F. Heinz, and D. A. Reis, High-harmonic generation from an atomically thin semiconductor, Nat. Phys. **13**, 262 (2017).
[58] F. Nastos and J. E. Sipe, Optical rectification and shift currents in GaAs and GaP response: Below and above the band gap, Phys. Rev. B **74**, 035201 (2006).
[59] L. Z. Tan and A. M. Rappe, Enhancement of the Bulk Photovoltaic Effect in Topological Insulators, Phys. Rev. Lett. **116**, 237402 (2016).
[60] O. Neufeld, N. Tancogne-Dejean, U. De Giovannini, H. Hübener, and A. Rubio, Light-Driven Extremely Nonlinear Bulk Photogalvanic Currents, Phys. Rev. Lett. **127**, 126601 (2021).
[61] A. Bharti and G. Dixit, Tailoring photocurrent in Weyl semimetals via intense laser irradiation, Phys. Rev. B **108**, L161113 (2023).
[62] D. Golde, T. Meier, and S. W. Koch, High harmonics generated in semiconductor nanostructures by the coupled dynamics of optical inter- and intraband excitations, Phys. Rev. B **77**, 075330 (2008).
[63] T. Higuchi, M. I. Stockman, and P. Hommelhoff, Strong-Field Perspective on High-Harmonic Radiation from Bulk Solids, Phys. Rev. Lett. **113**, 213901 (2014).
[64] P. G. Hawkins, M. Yu. Ivanov, and V. S. Yakovlev, Effect of multiple conduction bands on high-harmonic emission from dielectrics, Phys. Rev. A **91**, 013405 (2015).
[65] T. Ikemachi, Y. Shinohara, T. Sato, J. Yumoto, M. Kuwata-Gonokami, and K. L. Ishikawa, Trajectory analysis of high-order-harmonic generation from periodic crystals, Phys. Rev. A **95**, 043416 (2017).
[66] T. Ikemachi, Y. Shinohara, T. Sato, J. Yumoto, M. Kuwata-Gonokami, and K. L. Ishikawa, Time-dependent Hartree-Fock study of electron-hole interaction effects on high-order harmonic generation from periodic crystals, Phys. Rev. A **98**, 023415 (2018).
[67] T. Otobe, High-harmonic generation in $\ensuremath{\alpha}$-quartz by electron-hole recombination, Phys. Rev. B **94**, 235152 (2016).
[68] K. L. Ishikawa, Nonlinear optical response of graphene in time domain, Phys. Rev. B **82**, 201402 (2010).
[69] K. L. Ishikawa, Electronic response of graphene to an ultrashort intense terahertz radiation pulse, New J. Phys. **15**, 055021 (2013).
[70] Y. Morimoto, Y. Shinohara, K. L. Ishikawa, and P. Hommelhoff, Atomic real-space perspective of light-field-driven currents in graphene, New J. Phys. **24**, 033051 (2022).
[71] Z.-Y. Chen and R. Qin, Circularly polarized extreme ultraviolet high harmonic generation in graphene, Opt. Express **27**, 3761 (2019).
[72] D. Xiao, M.-C. Chang, and Q. Niu, Berry phase effects on electronic properties, Rev. Mod. Phys. **82**, 1959 (2010).
[73] M. Gradhand, D. V. Fedorov, F. Pientka, P. Zahn, I. Mertig, and B. L. Györffy, First-principle calculations of the Berry curvature of Bloch states for charge and spin transport of electrons, J. Phys. Condens. Matter **24**, 213202 (2012).
[74] X. Xu, W. Yao, D. Xiao, and T. F. Heinz, Spin and pseudospins in layered transition metal dichalcogenides, Nat. Phys. **10**, 343 (2014).
[75] L. Yue and M. B. Gaarde, Introduction to theory of high-harmonic generation in solids: tutorial, JOSA B **39**, 535 (2022).
[76] J. Wilhelm, P. Grössing, A. Seith, J. Crewse, M. Nitsch, L. Weigl, C. Schmid, and F. Evers, Semiconductor Bloch-equations formalism: Derivation and application to high-harmonic generation from Dirac fermions, Phys. Rev. B **103**, 125419 (2021).

[77] S. A. Sato, M. Lucchini, M. Volkov, F. Schlaepfer, L. Gallmann, U. Keller, and A. Rubio, Role of intraband transitions in photocarrier generation, Phys. Rev. B **98**, 035202 (2018).
[78] M. S. Okyay, S. A. Sato, K. W. Kim, B. Yan, H. Jin, and N. Park, Second harmonic Hall responses of insulators as a probe of Berry curvature dipole, Commun. Phys. **5**, 1 (2022).
[79] The supplementary information details the $k$-point sampling for analyzing excitations in the Dirac region. It comprises analyses of laser-induced excitation occupations, Semiconductor Bloch Equations calculations for current decomposition, analytical discussions, and TDDFT calculations of Hall harmonics for the MoS2 monolayer explaining the Berry curvature contributions, which includes Refs. [16-17, 19, 76–77] and additional Ref [80].
[80] I. Floss, C. Lemell, G. Wachter, V. Smejkal, S. A. Sato, X.-M. Tong, K. Yabana, and J. Burgdörfer, Ab initio multiscale simulation of high-order harmonic generation in solids, Phys. Rev. A 97, 011401 (2018).